\documentclass{iaus}
\usepackage{graphicx}

\title[SFH and CEH of the SMC] 
{Old main-sequence turnoff photometry in the SMC: Star Formation History and Chemical Enrichment Law}

\author[No\"el et al.]   
{Noelia E. D. No\"el$^1$, Carme Gallart$^1$,  Antonio Aparicio$^1$,
Sebasti\'an L. Hidalgo$^2$, Ricardo Carrera$^1$, 
 Edgardo Costa$^3$ \and Ren\'e A. M\'endez$^3$}

\affiliation{$^1$Instituto de Astrof\'\i sica de Canarias. 38200 La
Laguna. Tenerife, Canary Islands. Spain \break email: 
noelia@iac.es\\[\affilskip]$^2$University of Minnesota, 
Department of Astronomy, 
116 Church St. S.E., 
Minneapolis, MN 55455\\[\affilskip]
$^3$Departamento de Astronom\'\i a, Universidad de Chile, Casilla 36-D,
 Santiago, Chile}

\pubyear{2007}
\volume{241}  
\pagerange{119--126}
\date{?? and in revised form ??}
\jname{Stellar Populations as Building Blocks of Galaxies}
\editors{A. Vazdekis \& R. Peletier, eds.}
\begin{document}

\maketitle

\begin{abstract}
We present deep ground-based {\it B} and {\it R} observations of 12 fields in the Small Magellanic Cloud (SMC). 
The resulting color-magnitude diagrams (CMDs) reach the oldest main-sequence (MS) turnoff at M$_{R}$$\thicksim$3.5 and
reveal the stellar population differences between the part of the galaxy facing the Large Magellanic Cloud (LMC) and an 
area on the opposite side.  
In the Southern part of the galaxy, we found that there are still intermediate-age stars as far as 4 kpc from the SMC center. 
 The Chemical Enrichment History (CEH) in one of our SMC fields is also presented. 
 
\end{abstract}

\firstsection 

\begin{figure}
\begin{center}

\includegraphics[height=4.2in,width=4.2in]{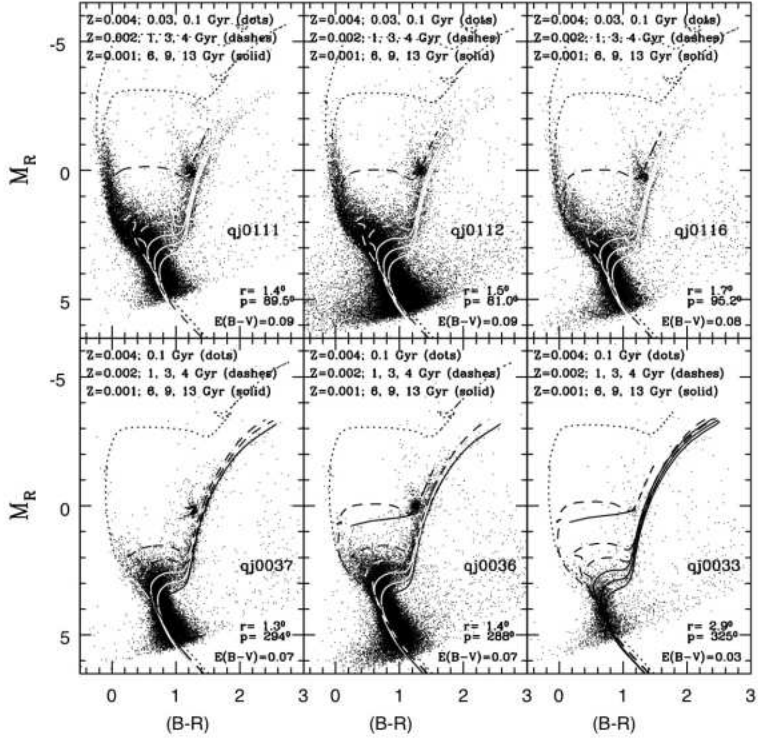}

\caption{Upper panel: CMDs of the Eastern SMC fields.
  Lower panel: CMDs of the Western fields. See text for details.}\label{fig:fig1}
\end{center}
\end{figure}

\begin{figure}
\begin{center}

\includegraphics[height=4.2in,width=4.2in]{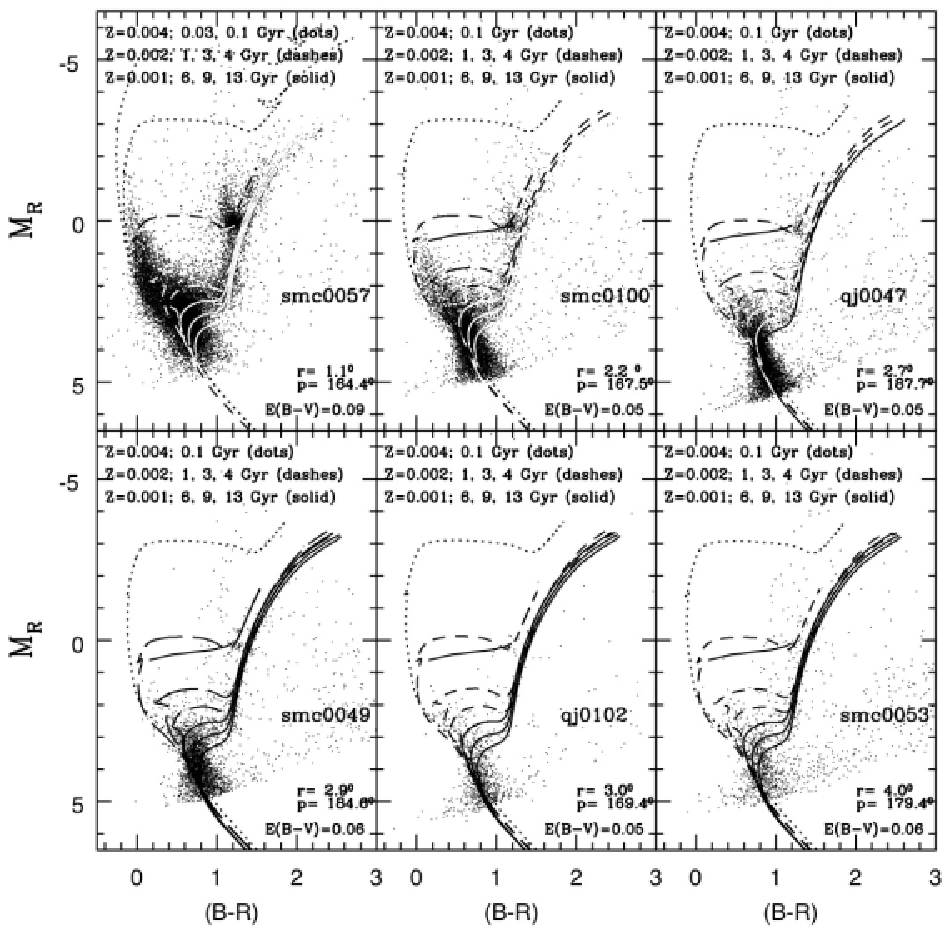}

\caption{CMDs of the six Southern SMC fields.}\label{fig:fig2}
\end{center}
\end{figure}

\section{Introduction}

In spite of the increasing interest in studying the SMC, reflected by recent works (e.g. Dolphin et al. 2001; 
Harris \& Zaritsky 2004), there are still many gaps in our knowledge of this galaxy.
 To shed light on some of these, we present
  a progress report of a project aimed at studying the star formation history (SFH)
 in different parts of the SMC. We observed 12 fields during a four years
campaign (2001-2004) using the 100-inch telescope at Las Campanas Observatory, Chile.
 The fields are located from $\thicksim$1 kpc to $\thicksim$4 kpc from the center
 of the SMC.
We will show a preliminary analysis of the age and metallicity ranges in each of the fields through isochrone's comparison, while a
quantitative analysis is underway through comparison with synthetic CMDs.
We also present the CEH of a SMC field, located
at $\thicksim$1 kpc from the center, as obtained from our synthetic CMD analysis. This CEH
is compared with the CEH found by Carrera (PhD thesis) for the same field 
using the CaII triplet, i.e., a completely independent method.

\section{The SMC Stellar Content}

The interpretation of CMDs of composite stellar population strongly relies on the stellar evolution models adopted (see
Gallart, Zoccali, \& Aparicio 2005). For our purpose we used   
the BaSTI stellar evolutionary models (Pietrinferni et al. 2004).
From the isochrones' overlap it is possible to constrain the range of metallicities in the SMC. However, the
method we use to calculate the SFH, involving comparison with synthetic CMDs (see No\"el et al. 2007, this conference),
allows us to obtain a reliable, more detailed CEH as seen in the next section.

Figures~\ref{fig:fig1} and~\ref{fig:fig2} show our SMC CMDs with isochrones overlapped. The CMDs of the different fields 
are displayed in order of increasing galactocentric distance.
Position angles (p), reddening E(B-V), metallicities and ages of the
isochrones are labeled in both figures. 
A distance modulus (m-M)$_{0}$=18.9 has been assumed.

The upper panel in Figure~\ref{fig:fig1} shows the Eastern SMC CMDs. Note
 the conspicuous MS, which is well populated from the oldest turnoff
at M$_{R}$=3.5 up to the 0.03 Gyr isochrone. All the CMDs show
 a large fraction of young stars ($<$1 Gyr old). The densely populated areas around the 
 isochrones of age $>$1 Gyr and the extension in luminosity of the red clump indicate a strong presence of 
intermediate-age stars.

The lower panel in Figure ~\ref{fig:fig1} shows the Western SMC CMDs. It is noticeable that the star formation
significatively dropped $\thicksim$3 Gyr ago (see No\"el et al. 2007).
Even in the fields at larger galactocentric distances, the population is not purely old but dominated by 
intermediate-age stars, as indicated by the relatively high density of stars around the 3, 4, 6, and 9 Gyr isochrones.

 Figure ~\ref{fig:fig2} shows the CMDs corresponding to the Southern fields. 
Field smc0057 is the closest to the SMC center. Its CMD shows an important MS, with the area around the 0.1 Gyr
 isochrone still quite populated. 
  The more distant fields toward the South are dominated by 
intermediate-age stars. These CMDs show 
a breaking point further than $\thicksim$2.7$^\mathrm{0}$, from where there are almost no stars younger than 3 Gyr old.

In all fields, the subgiant branch region is well populated from the oldest turnoffs up to 
the $\thicksim$1 Gyr
 old turnoff, indicating that star formation has proceeded in a continuous way, with possible variations in intensity but no
 big gaps between successive bursts, over the galaxy's lifetime. The structure of the red clump of core He-burning stars is consistent
 with the large amount of intermediate-age population inferred from the MS and the subgiant branch region. 
 The absence of a horizontal branch is a shared feature of all of the CMDs (see No\"el et al. 2007 for details).

 Our analysis shows that the
 underlying spheroidally distributed
population is mainly composed by intermediate-age and old stars, and that its age distribution does
not show strong galactocentric gradients.

\section{The Metallicity Law}

\begin{figure}
\begin{center}

\includegraphics[height=3in,width=3.5in]{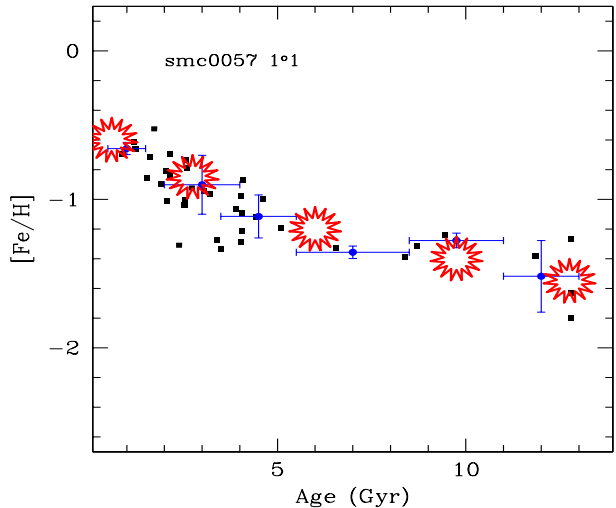}

\caption{CEH of field smc0057. Small (black) squares: measurements of the metallicity of individual stars by Carrera (PhD Thesis),
obtained from CaII triplet spectroscopy. Horizontal and vertical bars represent
  the metallicity dispersion corresponding to a given age range. Large stars (red symbols) represent our CEH
  obtained from the CMD modelling. Note the very good agreement of the two independent CEH.
}\label{fig:fig5}
\end{center}
\end{figure}

 As explained in detail in No\"el et al. (2007, this conference), we used the
 algorithm IAC-pop to solve the SFH of a SMC field (smc0057). The details of IAC-pop and its
 internal consistency are explained in Aparicio \& Hidalgo (2007, submitted). 
 See No\"el et al. (this conference) for a description of the process involved in the determination of the SFH of this particular field.
  The solution for the SFH provided by IAC-pop gives an independent determination of the CEH: 
  each age range has an associated metallicity distribution. We have calculated the mean metallicity of the stars formed at each age
  interval. We plotted this mean metallicity obtained from the CMD modelling in Figure~\ref{fig:fig5} (large stars),
   together with
  measurements of the metallicity of individual stars in the same field 
   (small black squares), using CaII triplet spectroscopy, from Carrera (PhD Thesis).
 Our CEH agrees very well with the one
 obtained by Carrera using a completely independent method. Thus, we have tested, for the first time, the 
 external consistency of the IAC-pop algorithm.
 It should be stressed that we did not impose any {\it a priori} constraints on the CEH. 

  We found that the mean metallicity of stars formed in this SMC field remained relatively low ([Fe/H]$\thicksim$-1.3)
   until $\thicksim$4 Gyr ago, at which point it began a steady increase
   to the present gas-phase abundance value of [Fe/H]$\thicksim$-0.5.
 The CEH found here
 is similar to the one found for clusters in the SMC (e.g. Pagel \& Tautvai$\check{s}$ien\'e 1999),
  with the special feature that there is only one cluster
 older than 10 Gyr (NGC 121, Suntzeff et al. 1986),  whereas in the field population there seems to be a significant fraction of
  old stars.

\end{document}